\pdfoutput=1

\documentclass[fleqn,11pt]{wlscirep}

\usepackage{lineno}
\usepackage{amssymb}
\usepackage{bm}
\usepackage{enumitem}   
\usepackage{graphicx}
\usepackage{fullpage}
\usepackage{setspace}
\usepackage{hyperref}
\usepackage[font=small]{caption}
\usepackage{subcaption}
\usepackage{tocloft}
\usepackage{longtable}  
\usepackage{multirow}  
\usepackage{verbatim}
\usepackage{totcount}
\usepackage{xcolor}

\usepackage{adjustbox}
\usepackage{orcidlink}

\usepackage{float} 

\graphicspath{{Figures/}}

\newcommand{\figref}[2]{Fig.\ \ref{#1}\textbf{#2}}

\newcommand{\revision}[1]{{\color{blue}{#1}}}

\begin{document}

\title{The increasing fragmentation of global science limits the diffusion of ideas}

\author[1,*]{Alexander J.\ Gates\orcidlink{0000-0003-0099-7480}}
\author[1]{Jianjian Gao\orcidlink{0009-0000-2436-1917}}
\author[2]{Indraneel Mane}

\affil[1]{School of Data Science, University of Virginia, Charlottesville, Virginia, USA}
\affil[2]{Network Science Institute, Northeastern University, Boston, Massachusetts, USA}
\affil[*]{To whom correspondence should be addressed: \href{mailto:agates@virginia.edu }{agates@virginia.edu}}




\begin{abstract}
\textbf{Abstract}\\

{\setstretch{1.5}
Global science is often portrayed as a unified system of shared knowledge and open exchange. 
Yet this vision contrasts with emerging evidence that scientific recognition is uneven and increasingly fragmented along regional and cultural lines.
Traditional models emphasize Western dominance in knowledge production but overlook regional dynamics, reinforcing a core-periphery narrative that sustains disparities and marginalizes less prominent countries.
In this study, we introduce a rank-based signed measure of national citation preferences, enabling the construction of a global recognition network that distinguishes over- and under-recognition between countries. 
Using a multinomial logistic link prediction model, we assess how economic, cultural, and scientific variables shape the presence and direction of national citation preferences. 
We uncover a global structure composed of multiple scientific communities, characterized by strong internal citation preferences and negative preferences between them---revealing growing fragmentation in the international scientific system. 
A separate weighted logistic regression framework suggests that this network significantly influences the international diffusion of scientific ideas, even after controlling for common covariates.
Together, these findings highlight the structural barriers to equitable recognition and underscore the importance of scientific community membership in shaping influence, offering valuable insights for policymakers aiming to foster inclusive and impactful global science.
}
\end{abstract}

\maketitle

\pagebreak


\setstretch{1.5}


\noindent
The global scientific ecosystem is shaped by the emergent interplay between international collaboration, competition, and recognition, which collectively shape the pathways, pace, and patterns of idea diffusion as well as how influence circulates globally~\cite{hagstrom_competition_1974, chinchilla-rodriguez_follow_2019, marginson2022drives}.
Strong national research infrastructures empower countries to vie for competitive advantages in technology, economics, security, and health.
Concurrently, scientific knowledge flows on a global scale, with scientific ideas---whether conceptual advances, methodological innovations, or technical breakthroughs---disseminating from their nation of origin and influencing research around the world. 
Yet, the strength of influence is far from uniform; nations vary widely in their visibility, the degree to which their work is recognized, and their capacity to shape global research agendas---producing persistent patterns of stratification and unequal recognition across the scientific landscape~\cite{moravcsik_applied_1985, schott_ties_1998, galvez_scientific_2000, tickner_core_2013, collyer_sociology_2014, gomez2022leading}.

Mapping the structure of global science provides critical insights into the dynamics of participation, influence, and recognition across countries~\cite{marginson2022drives}.
It helps uncover structural barriers that limit equitable engagement and reveals how scientific ideas move across national and disciplinary boundaries~\cite{wuestman2019geography}. 
Such analyses can expose patterns of dependency and bring attention to emerging scientific communities or regional clusters that are often overlooked in traditional rankings. 
These maps also help track how geopolitical shifts---such as U.S.-China tensions or evolving EU research policy---reshape global collaboration networks and redistribute scientific influence~\cite{zhou2006emergence}. 
Understanding these patterns is essential for designing national science strategies aimed at increasing international visibility and impact, whether through targeted investments, global partnerships, or improved language accessibility~\cite{scott2015dynamics}. 
Moreover, mapping recognition structures can strengthen science diplomacy by highlighting networks of trust and exchange that transcend borders. 
Perhaps most importantly, this work can help address structural imbalances in whose science shapes global priorities in fields like climate, health, and development. 
Yet despite this potential, most existing maps of global science remain focused on publication volume---obscuring more meaningful and nuanced relationships within the global research system~\cite{schott1988international}.

Here, we introduce a rank-based signed measure of national citation preference that quantifies the extent to which one country over- or under-recognizes the scientific work of another. 
Our approach offers a statistically grounded and equitable alternative to traditional measures based on raw citation or collaboration counts, which are often distorted by publication volume and fail to distinguish between meaningful preferences and structural noise.
Leveraging this measure, we construct a signed and directed international recognition network, which we analyze over time to uncover structural patterns of scientific recognition, including community formation, persistent asymmetries in recognition, and increasing international fragmentation. 
Using a multinomial logistic link prediction model, we further assess how economic, cultural, and scientific variables shape the presence and direction of national citation preferences.
Beyond structural insights, we demonstrate that the recognition network reflects substantive constraints on knowledge dissemination: using a logistic regression-based diffusion model, we show that positive citation preferences between countries significantly increase the probability that scientific ideas diffuse, while negative preferences suppress it.
The paper proceeds by reviewing the background on the structure of global science while highlighting the limitations of existing frameworks, introducing our methodology, and presenting empirical results on community formation, fragmentation, and idea diffusion. 
We conclude by discussing the implications of our findings for equitable scientific recognition, knowledge dissemination, and global science policy.

\subsection*{The structure of global science}
The prevailing theories for the structure and consequences of global scientific recognition closely mirror economic models, with a clear hierarchy and power dynamics between the ``core'' of scientific knowledge production and its ``periphery'' such that certain regions or countries dominate the production and dissemination of scientific research while others occupy a peripheral or marginalized position~\cite{prebisch1962economic, Shils1975center, may1997prodnations, king2004impactnations, zelnio_identifying_2012}. 
This core-periphery structure is hypothesized to have important consequences for international science by hindering diverse perspectives and the diffusion of ideas. 
The core-periphery model tends to oversimplify the complex relationships between nations, reducing influence dynamics to a binary classification of ``core'' or ``periphery'', while overlooking the nuances and inter-dependencies that shape global science~\cite{schott1988international}. 
By relying on this model, policy and funding decisions risk becoming skewed in favor of established centers, reinforcing existing national disparities. 
Core countries dominate research agendas and attract greater resources, while peripheral regions struggle to keep pace, further entrenching their marginal position in the global scientific network~\cite{sumathipala_underrepresentation_2004, kozlowski_intersectional_2022, abramo_role_2020, heimeriks_path_2014}.

Yet, it is often argued that the core-periphery model is entrenched in a Western-centric perspective that prioritizes resources and personnel, and thus overlooks the diverse cultural influences and research priorities shaping global scientific recognition and influence~\cite{schott_international_1988, seth_putting_2009, marginson2022drives}.
As early as 1988, Schott~\cite{schott_international_1988} suggested that the core-periphery structure is primarily attributed to the volume of a nations' scientific output which obfuscates the importance of other key factors related to ties between countries, such as geopolitical relationships, linguistic similarities, colleagueship, scientific cooperation, and educational connections.
Indeed, publication output remains heavily concentrated in the United States and a few European nations, implying that most quantitative indicators of scientific recognition—such as those based on raw publication, collaboration, and citation counts—tend to be notoriously Western-centric~\cite{gomez2022leading, may1997prodnations, king2004impactnations}. 
These metrics often overlook contributions from regions with smaller output, failing to recognize the diverse intellectual contributions and local innovations that may not fit neatly within dominant Western frameworks~\cite{anderson2018remembering}. 
These limitations highlight the need for more nuanced approaches that account for regional and contextual variations in scientific production and influence.

Recent observations challenge the longstanding Western-centric narrative, indicating that emerging scientific nations are reshaping the global landscape of scientific recognition. 
Countries such as China, Singapore, and South Korea are increasingly disrupting the traditional dominance of Western nations, signaling a shift in the concentration of global scientific influence~\cite{gui2019globalization, choi_core-periphery_2012, lariviere2018citationsstrength, basu_chinas_2018, leydesdorff2013international}.
However, this transformation has given rise to a growing tension between two perspectives. 
One emphasizes the rise of individual nations’ that have transitioned from the periphery to the core.
The other perspective critiques this vertical framing altogether, instead highlighting the emergence of regional alliances and calling attention to the persistent under-recognition of science from regions like Latin America, the Middle East, and East Asia---regions that remain structurally marginalized despite growing scientific capacity.
The latter perspective is best articulated by Marginson~\cite{marginson2022drives} who discusses ``the collapse of the centre-periphery model'' which he attributes to internal collaboration and regional alliances rather than through traditional engagement with Euro-American scientific hubs~\cite{marginson2023hegemony}.
Adams~\cite{adams2012rise} further characterizes such regional collaboration as a form of mutual recognition among partners within the region, fostering the development of emerging research economies.

These developments point to a broader structural transformation: rather than a linear shift where countries transition from periphery to core, the global scientific system may be evolving into a more decentralized and modular configuration, marked by strong internal ties within regions and weakening reliance on traditional Western hubs. 
This raises important questions about the cohesiveness of the global research system and whether its increasing regionalization reflects a fragmentation of scientific recognition. 
While modularity in collaboration and citation networks can signal healthy diversification, it may also indicate the formation of internally cohesive clusters with limited cross-regional exchange---an emerging pattern that aligns with concepts of fragmentation developed in the global governance literature~\cite{biermann2009fragmentation}.
Recent work by Greenhill and Lupu~\cite{greenhill2017clubs}, for example, defines fragmentation in international networks as the extent to which actors are organized into ``clubs'' with dense internal connections and sparse external ties. 
Similarly, Kim~\cite{kim2020global} situates fragmentation within a broader framework of polycentricity and complexity, emphasizing that fragmentation captures the disconnectedness across structurally distinct clusters, even in systems that are globally extensive. 
Applying this framework to global science would allows us to interrogate whether any observed modularity represents productive regional pluralism---or whether it reflects deeper structural barriers that constrain the flow of ideas across geopolitical divides. 
In this light, fragmentation becomes a critical concept for analyzing how regional alliances and shifting citation preferences are reshaping the global architecture of scientific recognition.

However, the tension between the core-periphery model and the emergence of regional alliances remains unresolved, largely due to a lack of robust quantitative evidence comparing the rise of individual countries within the existing core-periphery hierarchy with the creation of distinct regional scientific communities. 
Quantitative analyses are crucial for determining whether these regional networks are merely reinforcing the global hierarchy or truly reshaping it. 
Without data-driven comparisons, it remains unclear whether the traditional core-periphery model still applies or if a more nuanced framework is needed to capture the evolving dynamics of global scientific influence.

\subsection*{Mapping global science}

Efforts to map global science have drawn heavily on network-based approaches, capturing different dimensions of scientific activity through collaboration, citation, and mobility networks.
Collaboration networks are typically constructed from co-authorship data, linking researchers, institutions, metropolitan areas, or countries based on joint publication activity. 
These networks have revealed a growing tendency toward international collaboration, with certain countries---such as the United States, China, and select European nations---consistently occupying central, highly connected positions in the global system. 
Citation networks, by contrast, trace the flow of knowledge and scientific recognition via references in scientific publications. 
They are commonly used to assess influence or impact and to map the transmission of ideas across national or disciplinary boundaries.
Yet, citation patterns are shaped by disciplinary conventions, journal visibility, and linguistic proximity, complicating their interpretation. 
Finally, mobility networks track the movement of researchers by analyzing changes in author affiliations across time and publications. 
These networks are particularly valuable for understanding the global distribution of scientific talent, patterns of ``brain drain'' or ``brain circulation'', and the formation of transnational research communities.

Quantitative analysis of these networks frequently reveal evidence for the core-periphery structure of global scientific recognition.
For example, international collaboration networks show that core countries have higher degrees of centrality and connectivity than periphery countries, indicating their dominant role in global science~\cite{zelnio_identifying_2012, leydesdorff2008internationalcollab, gui2019globalization, choi_core-periphery_2012, wagner_continuing_2015}, and the global embeddedness of a nation, quantified by proportion of internationally co-authored publications, is a significant predictor of traditional scientific impact~\cite{wagner_open_2017}.
Additional analysis utilizing hierarchical clustering and dominant flow methodologies on international collaboration networks suggest that the global scientific community consists of four tiers: core, strong semi-periphery, semi-periphery, and periphery~\cite{gui2019globalization}.  
Under this model, the United States consistently occupies the core, maintaining collaborations with nearly every major scientific nation, while emerging powers like China and South Korea have only recently ascended to the core.
Mobility patterns also reveal that core countries attract more foreign scientists and researchers than periphery countries, suggesting their greater availability of resources and opportunities~\cite{freeman_globalization_2010, scott2015dynamics, adams1998benchmarking, urbinati2021measuring, bauder2018internationally}. 
Scott~\cite{scott2015dynamics} refers to this phenomenon as ``hegemonic internationalisation'' where internationalization becomes an extension of global inequality and the struggle for dominance, driven by competition, rankings, and the concentration of academic power in certain geopolitical centers.
Analysis of raw citation networks further demonstrate that core countries generate more citations than periphery countries, implying their higher impact and influence on scientific research~\cite{schott_ties_1998, gomez2022leading, schott_international_1988, choi_core-periphery_2012}. 
Notably, Gomez et al.~\cite{gomez2022leading} draws on the existing classification of countries into core and periphery to reveal a growing disparity between the number of citations a country receives and the textual similarity of the publications they produce~\cite{gomez2022leading}.

Together, the quantitative analysis consistently appears to converge on a core-periphery model that centralizes a small group of scientifically dominant countries.
However, most of these results rely heavily on raw counting methods---simple tallies of co-authorships, citations, or researcher movements---which are deeply susceptible to the volume problem~\cite{mingers2015review}.
The volume problem arises when countries or institutions with large publication outputs appear disproportionately central or influential---not due to stronger or more reciprocal scientific relationships, but simply because of their scale---thereby obscuring important dynamics within the global scientific system~\cite{leydesdorff1988problems}. 
First, it masks meaningful instances of under-recognition: if a country consistently produces research that is selectively overlooked in citation networks, raw counts will likely miss this asymmetry.
Second, it dilutes the signal of preferential interactions by privileging connections with high-volume countries. 
As a result, rich and substantive collaborations between smaller or less prolific countries are often rendered invisible, buried beneath the statistical weight of interactions with dominant players. 
The structure that emerges from raw counts, then, may reflect more about publication volume than about actual patterns of influence, preference, or mutual engagement.
This tendency is well captured by gravity models, which have been applied to global citation and collaboration networks to show that interaction volume between countries is largely predictable based on their output size and geographic distance~\cite{pan2012world}. 
These models reveal that much of what appears as scientific centrality or dominance can be explained by volume and proximity alone, rather than any intrinsic scientific preference or influence.
To further illustrate this distortion, we provide a calculation in the Supplementary Information (SI, Section S6.4) comparing network measures computed on observed versus randomized data.
The high correlation between metrics in the true and randomized networks suggests that much of the observed structure is volume-driven rather than preference-driven. 
These findings underscore the need for more refined approaches that account for statistical expectations and enable us to identify meaningful deviations, such as over- or under-recognition—relative to country-specific baselines.

Recognizing the limitations of raw counting and the distortions introduced by publication volume, several more nuanced approaches have emerged. 
One class of methods employs diffusion models which attempt to trace the flow of scientific recognition by simulating how influence spreads across the global citation network~\cite{zhang2013characterizing}. 
While these models remain influenced by volume, they can elevate countries that are significantly cited by major players, even if their overall output is low.
A second class of methods addresses the volume problem more directly by normalizing counts relative to a null model. 
These models typically define an expected citation or collaboration rate based on structural baselines---such as total output or field distribution---and then evaluate deviations from that baseline. 
For example, in Leydesdorff and Wagner~\cite{leydesdorff2008internationalcollab}, the authors apply cosine normalization to a country-by-country co-authorship matrix, revealing latent collaboration structures that are hidden in unadjusted data. 
Without normalization, the United States appears central in every respect, but after normalization, a more differentiated core emerges. 
Similarly, Schubert and Glanzel~\cite{schubert2006cross} proposed a normalized measure of co-authorship or citation affinity that compares observed counts to expected counts under random pairing, allowing for the identification of overrepresented or underrepresented relationships.
These approaches shift the focus from raw activity to preferential patterns of interaction, enabling finer-grained insights into the structure of global science and its inequalities.

While normalization methods represent an important advance over raw counting, they are not without limitations. 
First, many approaches depend on arbitrary thresholds, including what qualifies as a ``significant'' deviation from a null model or what proportion of citations is deemed indicative of a meaningful structural relationship. 
These thresholds can be sensitive to model assumptions, data sparsity, or temporal aggregation choices, and may not have a clear unit of interpretation. 
Second, although normalization reduces the influence of volume, it does not fully eliminate it. 
Larger countries still tend to dominate the normalized landscape, especially when null models assume proportional citation or collaboration rates, which can reintroduce the bias they aim to correct~\cite{vaccario2017quantifying}. 
Third, the interpretability of normalized metrics can be challenging---cosine similarity scores, for example, are sensitive to low-frequency interactions and may conflate structural alignment with actual preference or influence~\cite{leydesdorff2008normalization}. 
Finally, some normalization schemes rely on highly stylized null models (e.g., assuming random mixing), which may not reflect the true underlying dynamics of the global scientific system, such as linguistic or historical ties~\cite{leydesdorff2008normalization}. 
As a result, while these methods offer improved fairness over raw counts, they must be applied with caution—and ideally, complemented by statistical significance testing to distinguish noise from meaningful structural signals.

\subsection*{Scientific ideas, knowledge diffusion, and global recognition}

Understanding the structure of global science is essential for mapping the diffusion of scientific ideas because the pathways through which ideas spread are shaped not only by intellectual content but also by the social structures within which they emerge---the institutional, geographic, and cultural networks in which science is embedded~\cite{crane1972invisible, freeman2017migration, sugimoto2017scientists}. 
The organization of international scientific recognition---who cites whom, and which nations preferentially acknowledge others---plays a critical role in either facilitating or impeding the cross-border movement of novel concepts.
In bibliometric studies, scientific ideas are often conceptualized as identifiable terms or phrases (e.g., n-grams) that appear in the titles and abstracts of publications and can be traced over time and space~\cite{milojevic_quantifying_2015, frank2019evolutionai, hofstra2020diversity, cheng_how_2023, gao2024quantifyingai}. 
These linguistic proxies allow researchers to observe when and where an idea first emerges and how it diffuses to new research communities or national contexts. 
However, idea diffusion is not uniform; it is influenced by structural features such as citation preferences, collaboration patterns, and linguistic or geopolitical proximity~\cite{freeman_globalization_2010, heimeriks_path_2014}. 
By analyzing the architecture of global science---including who receives recognition and who is systematically overlooked---we gain a more complete understanding of how knowledge moves, who benefits from it, and where potential barriers to equitable dissemination lie.

Scientific recognition is not only a marker of scholarly influence---it is a gatekeeping mechanism that shapes which ideas gain visibility, credibility, and global reach~\cite{siler2015gatekeeping}. 
In a highly stratified research system, recognition tends to accumulate in already well-resourced countries and institutions, reinforcing existing hierarchies and privileging certain narratives, methodologies, and research priorities~\cite{miao2024persistenthier}. 
As a result, ideas emerging from less central or lower-resourced regions may struggle to diffuse internationally, regardless of their intrinsic merit or local relevance. 
This has direct implications for equity in global science, as it influences whose knowledge is amplified, whose voices are legitimized, and which problems receive attention on the global research agenda~\cite{kozlowski_intersectional_2022}. 
Understanding the relationship between recognition and diffusion thus allows us to assess not just how ideas spread, but whose ideas are allowed to spread, and why~\cite{cheng_how_2023}. 
These insights are critical for designing more inclusive research policies, supporting underrepresented regions, and fostering a genuinely global scientific community that reflects a diversity of contributions and perspectives.

To move beyond volume-distorted metrics and capture meaningful patterns of international scientific recognition, we propose a new method for quantifying national citation preferences. 
Our approach draws on rank-based comparisons between the distribution of citations from a source country and the global distribution of its citation activity in a given year.
This enables us to estimate whether a country is over- or under-recognized by another, independent of overall output. 
Importantly, this measure accommodates statistical significance testing and is robust to country size, allowing for a more nuanced and equitable mapping of global recognition. 
By embedding these relationships in a signed, temporal network, we examine the evolving topology of international scientific recognition and its implications for the flow of ideas, the formation of regional scientific communities, and the fragmentation of global science.

\revision{\section*{Results}}

\subsection*{A pairwise measure of national citation preference}

\begin{figure}[H]
    \centering
    \includegraphics[width=0.71\textwidth]{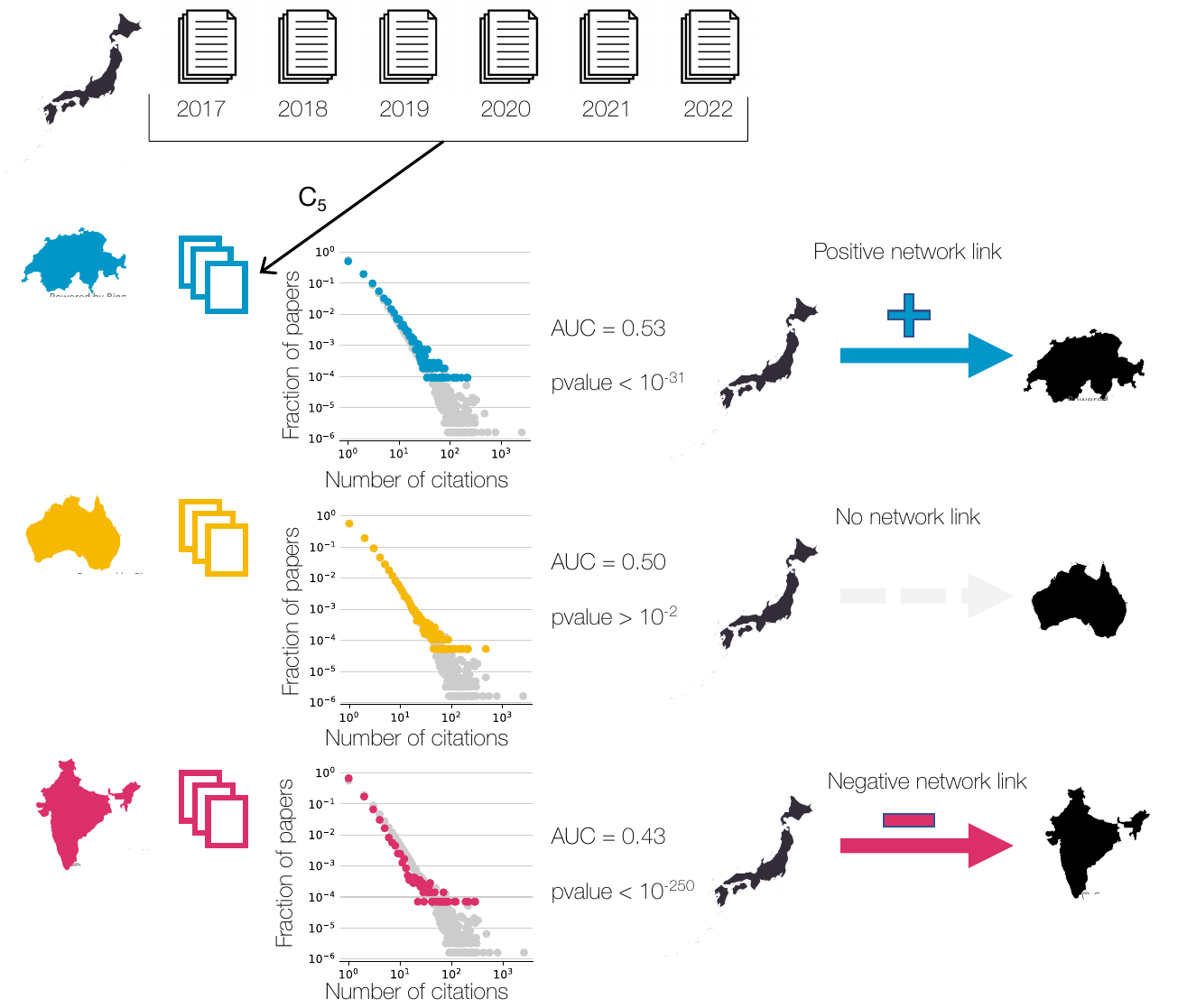}
     
    \caption{\textbf{Illustration of the measure of national citation preference}. 
    For articles published in 2017 in a target country (e.g. Switzerland, Australia, India), we take all citations within the next 5 years, $c_5$, from a source country (e.g., Japan, black).
    We compare the citation distribution to each target country (e.g., Switzerland, blue) against the citation distribution to all countries (grey), calculating the Area Under the Curve (AUC), equivalent to the Mann-Whitney U statistic, of which the statistical significance can be accessed.
    Each directed relationship is classified as positive (over-citation, e.g., Japan to Switzerland), statistically insignificant (no preference, e.g., Japan to Australia), or negative (under-citation, e.g., Japan to India).
    }
    \label{fig:measure}
\end{figure}

To quantitatively capture patterns of national scientific recognition, we adopt a probabilistic, rank based measure that assesses whether one country systematically over- or under-cites the scientific output of another. 
As illustrated in \figref{fig:measure}{}, our method compares the rank distribution of citations from a source country to a target country against the source country’s citations to the global publication landscape in the same year.
This yields a measure equivalent to the Area Under the Curve (AUC) in a two-sample rank test~\cite{conover1999practical} (Methods and SI, Section S3). 
It represents the probability that a randomly selected paper from the target country is cited more frequently by the source country than a randomly selected paper from the rest of the world.
A value of 0.5 indicates no preference, while values closer to 1 indicate strong over-citation (positive preference), and values closer to 0 indicate under-citation (negative preference). 
This AUC-based interpretation is also mathematically equivalent to the Mann-Whitney U statistic (also known as the Wilcoxon rank-sum statistic), allowing us to assess the statistical significance of over- or under-recognition without assuming normality in citation distributions.
To construct the appropriate baseline, we aggregate the source country’s outbound citations across a 5-year window for each publication year, smoothing short-term fluctuations while remaining sensitive to evolving patterns of international recognition.

\subsection*{International network of scientific recognition}

\begin{figure}[H]
    \centering
    \includegraphics[width=\textwidth]{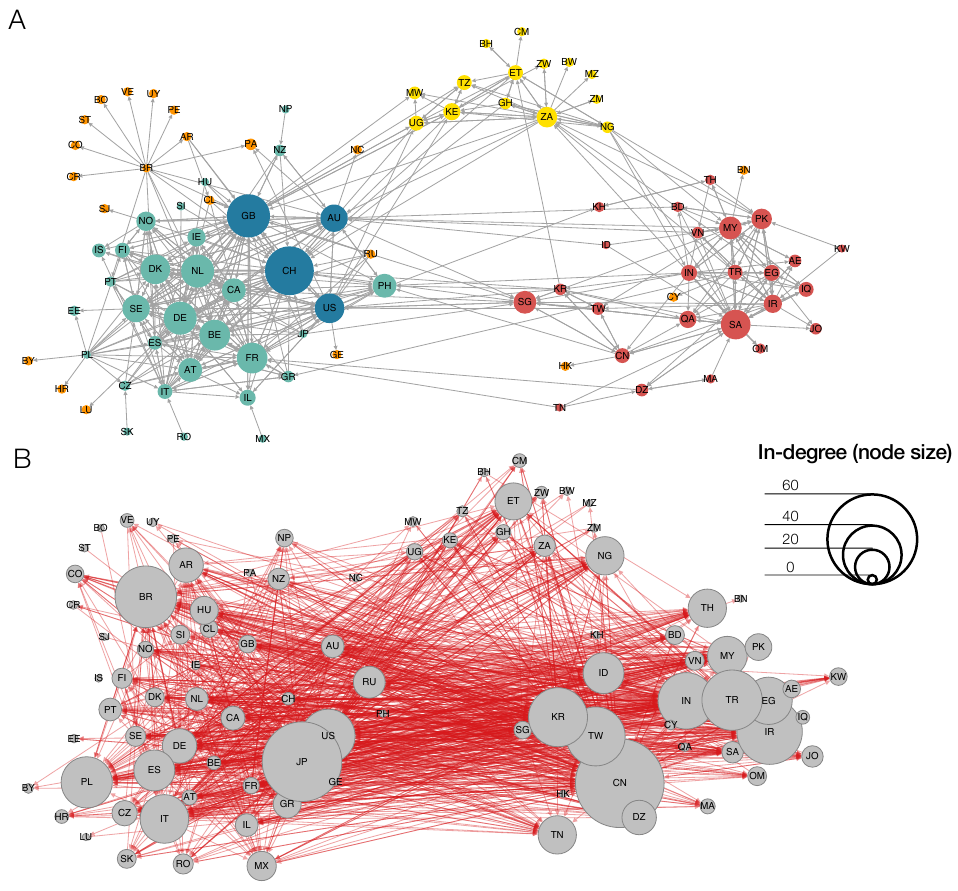}
    \caption{\textbf{International network of scientific citation preferences}.
    \textbf{A}) The positive citation preferences and B) negative citation preferences in the cumulative network.
    The node area captures the country in-degree, while node color reflects membership in one of five communities inferred using the degree-corrected stochastic block model.
    Node position is the same in both panels and was derived using only the positive relationships.
    }
    \label{fig:network}
\end{figure}

\begin{figure}[H]
    \centering
    \includegraphics[width=0.9\textwidth]{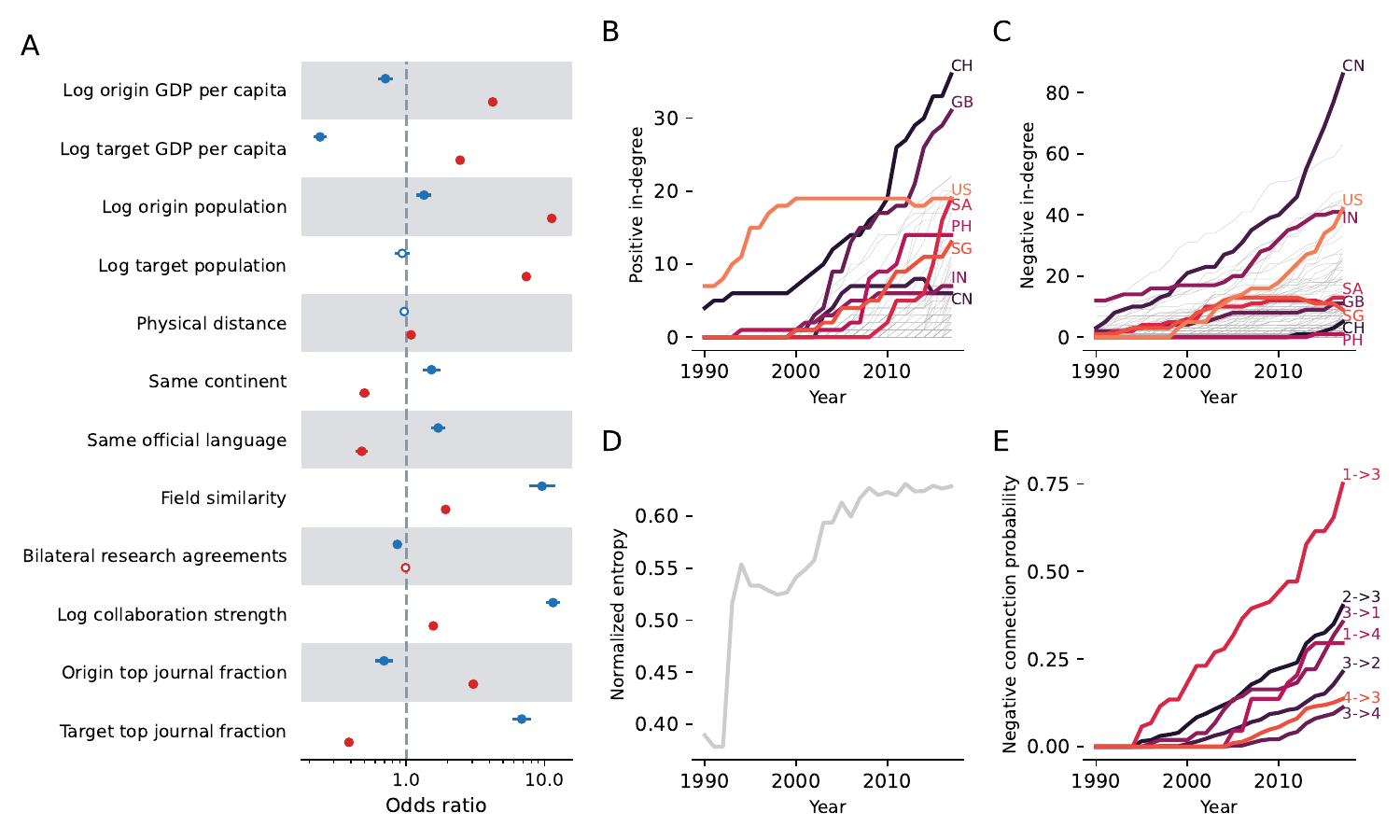}
    \caption{\textbf{Properties of the international network of scientific citation preferences}.
    \textbf{(A)} The odds ratios for a multinomial logit model with temporal fixed-effects to predict the positive (blue) or negative (red) citation preference compared to the baseline of no preference.  Solid points are statistically significant at $p<0.05$ with the 95\% confidence intervals shown.  The full regression table can be found in the SI, Table S3.
    The \textbf{(B)} positive and \textbf{(C)} negative in-degrees highlight 6 prominent countries, including the most positively viewed country in 2017, Switzerland (CH), and the most negatively viewed country, China (CN).
    \textbf{(D)} The normalized entropy for the distribution of PageRank centrality over the nodes has been increasing over the last 30 years.
    \textbf{(E)} The probability for a negative citation preference between a country in a source community and a country in a target community.
    }
    \label{fig:networkprop}
\end{figure}

We next build the network of international scientific recognition.
The international scientific recognition network is a temporal signed and directed network in which each country is a node, and a source country is linked to a target country by a positive (negative) edge if the source country over-cites (under-cites) the target country's publications.
To begin, we consider the cumulative network in which we aggregate over time, taking any edge that appears at least once throughout the 27 years, and shown in \figref{fig:network}{}.
We find that 147 countries had at least one statistically significant relationship to be included in the network.
Of the 21,462 possible international relationships, only 536 are positive interactions and 1471 are negative interactions.
Scientific publications from Switzerland are over-cited by the most other countries, with 36 incoming edges, followed by Great Britain, Germany, and the Netherlands (\figref{fig:networkprop}{B}).
On the other hand, publications from China are the most under-cited, with 86 incoming under-citation edges, followed by Japan, Iran and India (\figref{fig:networkprop}{C}).
We find that links in the citation preference network are highly persistent---fewer than 2\% of directed edges ever change sign over time---and that once a link emerges, it typically remains in the network, aside from minor fluctuations driven by changes in publication volume. 
This stability justifies our use of a cumulative representation that captures long-term patterns in scientific recognition while filtering out short-term noise (SI, Section S6.1). 
At the same time, the strength of these links exhibits interesting temporal patterns (e.g., SI, Figure S1), which we view as a promising direction for future work.
In this study, we focus on a categorical, sign-based approach---distinguishing over- and under-recognition---because it offers a robust and interpretable summary of persistent structural asymmetries in international citation behavior, while still yielding meaningful insights into the organization of global science.

To identify key country-specific and dyadic factors related to national citation preferences, we estimate a multinomial logit model with temporal fixed-effects to predict the citation preference between all directed country pairs from 1990 through 2017. 
Each directed country pair is classified into one of three mutually exclusive categories: positive preference, negative preference, or no significant preference. 
This categorical approach aligns with signed link prediction and treats all statistically non-significant relationships as equivalent, preventing over-interpretation of minor citation differences based on raw magnitudes (see Methods and SI, Section S6.4).
To verify the robustness of our model, we conducted a Variance Inflation Factor (VIF) analysis and found that all values fall between 1.0 and 3.0---well below the conventional threshold of 5---indicating low multicollinearity minimal risk of distortion in coefficient estimates (SI, Figure S9).

We estimate a sequence of six model specifications, progressively incorporating additional covariates: from basic economic and to geography, shared language, topical distance, bilateral science and technology agreements, and indicators of research volume and quality (SI, Table S6). 
The McFadden pseudo-$R^2$ ranges from $0.561$ to $0.639$.
While most predictors are statistically significant, many do not differentiate between positive and negative preferences, suggesting they primarily affect whether a citation preference exists rather than its direction (\figref{fig:networkprop}{A} and SI, Table S6).
For example, collaboration strength increases the likelihood of a preference but does not predict whether it is positive or negative, while topical similarity is predictive of positive preferences only.
However, three cultural indicators: common language ( $\beta_{positive} = 0.53$, 95\% $CI = [0.41, 0.65]$; $\beta_{negative} = -0.74$, 95\% $CI = [-0.84, -0.63]$), same continent ($\beta_{positive} = 0.42$, 95\% $CI = [0.27, 0.57]$; $\beta_{negative} = -0.69$, 95\% $CI = [-0.78, -0.6] $), and participation in Science and Technology Agreements (bilateral research agreement, $\beta_{positive} = -0.17$, 95\% $CI = [-0.19, -0.15]$; $\beta_{negative} = -0.01$, 95\% $CI = [-0.02, 0.0]$), relate to both the presence and sign of the national citation preference (\figref{fig:networkprop}{A}).
We also find that research quality, proxied by the share of publications in top journals (see Methods and SI, Section S4.2), strongly predicts citation recognition.
Countries with higher top-journal output are more likely to receive positive preferences from others ($\beta_{positive} = 1.51$, 95\% $CI = [1.39, 1.63]$), while countries that publish more heavily in lower-ranked venues tend to receive negative ones ($\beta_{negative} = -0.75$, 95\% $CI = [-0.8, -0.69]$).
Interestingly, these high-status countries are also less likely to assign positive preferences to others $\beta_{positive} = -0.29$, 95\% $CI = [-0.4, -0.18]$) and more likely to assign negative ones ($\beta_{negative} = 0.88$, 95\% $CI = [0.82, 0.94]$)---suggesting that scientific recognition follows a hierarchical structure, where influence flows upward more readily than downward, reinforcing stratification in the global system.

To validate these findings, we conducted a robustness check by re-estimating the model using the cumulative network till 2017.
This isolates cross-country variation from within-country temporal fluctuations and confirms that our core results are not dependent on panel structure. 
The main conclusions hold (SI, Table S8), supporting the interpretation that our results reflect stable cross-sectional relationships rather than artifacts of within-unit variation.
In addition, we applied 5-fold stratified cross-validation to evaluate model performance, preserving the class distribution across folds, demonstrating consistent model performance and further reinforcing the robustness of our results (SI, Table S7).

Together, these findings suggest that national citation preferences reflect a complex interplay between scientific quality, cultural proximity, and science diplomacy, rather than being reducible to size or collaboration volume alone.

Mapping the network of international preferences over time reveals the changing landscape of scientific diplomacy.
Specifically, the network of international citation preferences has evolved away from a core-periphery structure dominated by a few hubs to a more distributed structure, a change which we measure by the increasing normalized entropy for the distribution of normalized PageRank centrality scores (\figref{fig:networkprop}{D}).
For example, before 2000, the network was dominated by the United States, with relatively little positive scientific recognition of countries in Asia or Africa (\figref{fig:networkprop}{B}).
However, by 2010, Switzerland and Great Britain surpassed the United States in global recognition, and there were notable rises in recognition to Saudi Arabia, the Philippines, and Singapore (\figref{fig:networkprop}{B}). 
Throughout this period, China and Japan remained the most under-cited, dominating the negative citation preference network (\figref{fig:networkprop}{C}).

\section*{Growing international scientific fragmentation}

The preference of some nations for the scientific work of others, combined with the proliferation of negative biases against groups of countries, is a characteristic hallmark of international scientific fragmentation\cite{biermann2009fragmentation, aref2020balance}.
This pattern in citation patterns can stem from various factors, such as disciplinary biases, prevailing research trends, language barriers, geographical disparities, or ideological preferences. 
As a result, scientific fragmentation can distort the perception of the importance and impact of research, reinforce existing knowledge gaps, and impede the equitable dissemination and recognition of diverse scientific contributions.

To measure the dynamics of international scientific fragmentation, we first detect the presence of international communities using the degree-corrected stochastic block model, finding strong evidence for a partition of the positive network in 5 distinct communities (SI, Section S6).
Three blocks strongly resemble a three-layer core-periphery structure~\cite{gallagher_clarified_2021}.
Specifically, as visually apparent in \figref{fig:network}{A}, we identify a dense inner core of Western countries---US, GB, CH, AU---that consistently prefer each other's work (Community 1, dark blue), alongside a weaker secondary core composed of many European countries (Community 2, teal).
Countries in the stronger core are less consistent in their recognition of the weaker core.
In contrast, countries in the periphery (Community 5, orange) tend to be agnostic toward one another, yet display positive citation preferences toward both the strong and weak cores.

At the same time, this analysis confirms that the core-periphery structure is an oversimplification of the diverse communities in global science.
The international scientific recognition network reveals two additional communities outside of the Western scientific world: one community captures an international community predominately composed of Asian countries (3, red), including both East Asia and the Middle East, while another reflects the African nations (4, yellow).

The fragmentation of global science is evidenced by the distribution pattern of positive and negative citation preferences across scientific communities.
Overall, only 34\% of positive citation preferences occur between nations from different communities, whereas negative citation preferences predominantly cross community boundaries, with over 86\% occurring between nations from different communities.

The structure of the international citation preference network and its communities provides a more nuanced view of the differing roles nations play in shaping global scientific recognition and knowledge dissemination.
For example, while both Singapore and China have gained recognition for their scientific contributions\cite{zhou2006china}, our analysis shows that Singapore occupies a unique bridging role between different communities, whereas China, despite its prominence, remains within the Asian community without holding a central core position (see SI, Section S6.4 for a detailed analysis). 
Notably, our work highlights Saudi Arabia, Turkey, and Iran as occupying more central roles within the Asian scientific community.
Similarly, South Africa (ZA) stands out as a central node within the African scientific community, while the network reveals the distinct roles of Uganda and Nigeria as key bridges—Uganda connecting to Western communities and Nigeria to the Asian community.

To assess the dynamics of international scientific fragmentation, we look at the probability of forming negative or positive links.
Overall, we observe a growing tendency for nations to negatively judge the work of other nations as evidenced by the increase in negative connection probabilities (SI, Figure S4).
However, the community structure of the international scientific recognition network reveals that these preferences are not evenly distributed and are not primarily directed at specific nations. 
Instead, the fragmentation of global science appears to be influenced by the detected geopolitical communities.
As shown in \figref{fig:networkprop}{E}, the probability of inter-community negative preference links has grown significantly since 1990.  
The probability of negative inter-community links is largest between the Western and Asian communities, specifically communities $1\rightarrow3$ and $3\rightarrow1$ as well as $2\rightarrow3$ and $3\rightarrow2$, but has also significantly grown between the African and Western communities $1\rightarrow4$, $4\rightarrow1$ and the African and Asian communities $3\rightarrow4$, $4\rightarrow3$.
Significantly, there are nearly symmetric negative inter-community link probabilities, indicating the true fragmentation of the global scientific landscape into distinct communities cannot be explained by a core-periphery model.

\section*{Controlling for disciplinary and journal-level effects}
\label{sec:journal_impact}

We now extend our analysis by introducing additional controls to further explore factors influencing citation preferences.
Our framework seamlessly integrates a non-parametric approach that accounts for the field or journal in which each article is published, allowing us to control for variability in citation practices across disciplines and venues. 
By incorporating these controls and juxtaposing the new network against our original, this enhanced model provides a more refined understanding of how disciplinary and journal-specific effects interact with national-level citation behaviors, offering deeper insights into the structure of global scientific recognition.

Instead of relying on the full citation distribution for all publications cited by the source country, we construct a new baseline citation distribution using a stratified bootstrap approach that accounts for journal frequency (see Methods for details). 
This technique samples from the source country’s conditional citation distribution while ensuring the sampled set reflects the observed publication counts for each journal. 
By controlling for journal-level citation patterns--commonly used as proxies for scientific discipline and ``quality''--this method provides a more detailed benchmark, isolating national citation preferences from journal-related con-founders.

\begin{figure}[H]
    \centering
    \includegraphics[width=\textwidth]{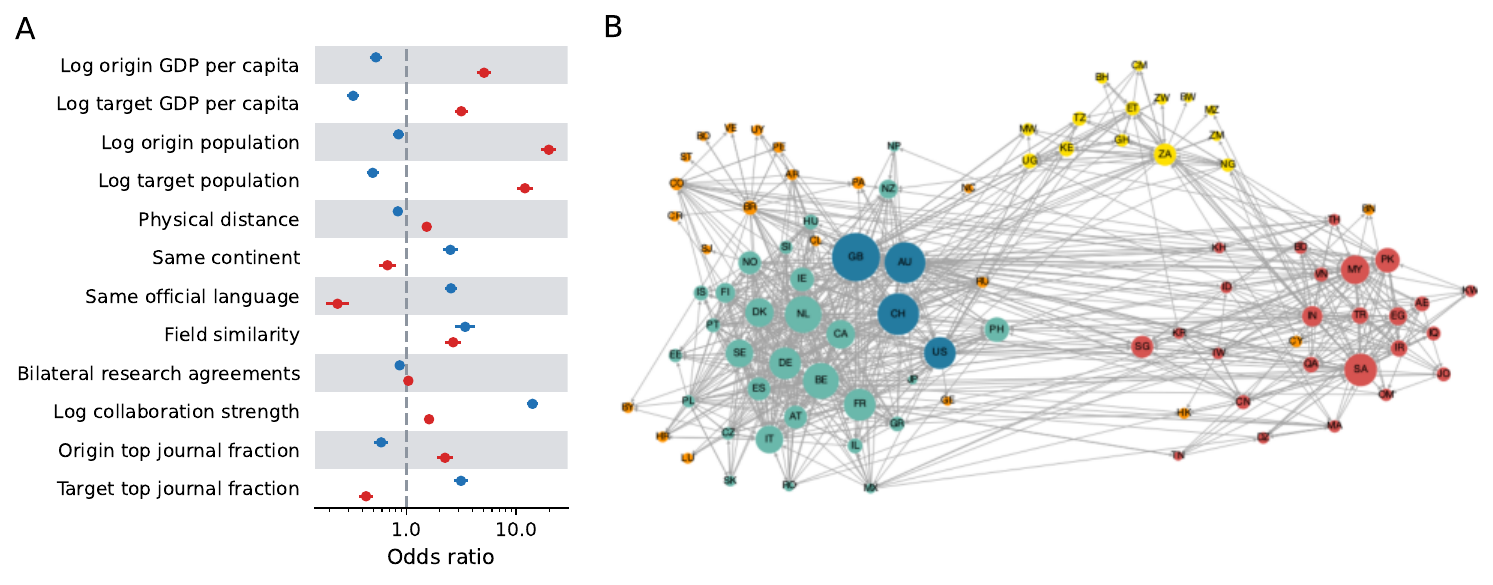}
    \caption{\textbf{The international network of scientific citation preferences controlling for publication journal}.
    \textbf{(A)} The odds ratios for a multinomial logit model to predict the positive (blue) or negative (red) citation preference compared to the baseline of no preference.  Solid points are statistically significant at $p<0.05$ with the 95\% confidence intervals shown.  The full regression table can be found in the SI, Table S4.
    \textbf{(B)} The journal bootstrap network filtered to positive relationships using the same layout as in Fig 1.
    }
    \label{fig:bootstrap}
\end{figure}

Shown in \figref{fig:bootstrap}{B}, the resulting cumulative international network of citation preferences based on the journal bootstrap (N2) exhibits both notable similarities and differences when compared to the original network (N1). 
Specifically, N2 reveals more positive national preferences, with a total of 645 compared to 541 in N1, while it shows significantly fewer negative preferences, dropping from 1,538 in N1 to just 334 in N2. 
At the same time, there is considerable overlap between the networks: 448 positive preferences are present in both networks, accounting for 84\% of the smaller N2, and 326 negative preferences are shared, representing 98\% of the smaller N1. 
The variation in positive edges is largely concentrated in a small number of countries: 47\% of the new edges are directed toward just 11 countries, while 30\% originate from only 7 countries.
Moreover, the edge distribution in N2 largely mirrors the community structure observed in N1 such that 60\% of positive edges connect nations within the same community in N2, slightly down from 66\% in N1, and 85\% of negative edges link nations from different communities in N2, compared to 86\% in N1.
Using a similar multinomial logistic regression model with temperal fixed-effects to predict the presence and sign of national preferences, we find the same independent variables play remarkably similar patterns of importance for predicting the odds of a positive or negative edge, and differentiating between those signs (\figref{fig:bootstrap}{A}).

Taken together, these observations suggest that about 80\% of the negative citation preferences we initially identified can be attributed to disciplinary differences in scientific focus and journal ``quality''.
However, we choose to retain these negative preferences in the original network because they reflect realized patterns of recognition---regardless of their disciplinary origin---and contribute meaningfully to the structural dynamics and fragmentation observed in global science.
At the same time, the increase in positive preferences primarily within the original communities indicates the importance of those communities, suggesting they are highly influential in shaping collaborative networks and recognition.
Ultimately, these findings emphasize the value of applying robust methodological frameworks to uncover the complexities of international citation preferences, providing deeper insights into the factors that influence scientific recognition on a global scale.

\section*{International recognition network limits the diffusion of ideas}

To explore the potential connection between the position of nations in the international scientific recognition network and the propensity for them to spread ideas, we investigate the diffusion of ideas between countries. 
We operationalize scientific ideas through the appearance of keywords in the title and abstract of scientific publications~\cite{milojevic_cognitive_2011, milojevic_quantifying_2015, cheng_how_2023}.
Specifically, we identify the mention of over 40,000 n-grams defined as scientific ideas by a previous study~\cite{cheng_how_2023} and limit to 22,413 unique ideas originating in only one country after 1990 (see Methods and SI, Section S5). 
We then model the probability that an idea originating in one country is eventually mentioned in another target country at least once during the subsequent 32 years (1990-2022) using a weighted logistic regression model. 
The dependent variable is the observed fraction of ideas from the origin country that are mentioned by an article in the destination country, and the weights correspond to the total number of ideas produced by the origin country, ensuring that observations are scaled by their underlying exposure.
This approach allows us to gauge the spread of information through the global scientific ecosystem, reflecting the broader exchange of ideas without needing to follow each idea's trajectory over time.
Consequently, we use the cumulative international recognition network where we aggregate into a static snapshot using all links that appear in at least one time slice.

To ensure the integrity of the regression model, we first assess multicollinearity among the independent variables using a Variance Inflation Factor (VIF) analysis. 
All VIF values fall well below the conventional threshold of 5, indicating no severe multicollinearity and confirming that the predictors contribute independently to the model (see SI, Section S9).

We begin by evaluating two baseline models that include standard control variables commonly used in cross-national diffusion studies. 
As shown in Table \ref{table:idea_diffusion}, both models (1) and (2) perform well, offering consistent and interpretable results that validate key structural and contextual predictors of idea diffusion.
Notably, both models achieve high explanatory power, with McFadden’s pseudo-$R^2$ values of $0.41$ and $0.42$, respectively---indicating that these baseline variables account for a substantial portion of the variation in diffusion outcomes.

Starting from Model (1), we find that economic prosperity in the origin country---measured by log GDP per capita---is positively associated with the international diffusion of scientific ideas ($\beta = 0.35$, 95\% $CI = [0.34, 0.35]$, odds ratio $\approx 1.41$), while GDP per capita in the destination country has a smaller positive effect ($\beta = 0.19$, 95\% $CI = [0.18, 0.19]$, odds ratio $\approx 1.21$), indicating that wealth in the origin country plays a larger role in enabling diffusion than in facilitating absorption.
Scholarly productivity in the destination country---measured by log publication volume---is the strongest predictor among these variables ($\beta = 1.66$, 95\% $CI = [1.65, 1.66]$, odds ratio $\approx 5.24$), likely because greater research output increases the number of opportunities to engage with and adopt ideas. 
In contrast, higher publication volume in the origin country is actually associated with a lower likelihood of diffusion ($\beta = -0.36$, 95\% $CI = [-0.36, -0.35]$, odds ratio $\approx 0.70$), suggesting that simply producing more research does not necessarily lead to wider international dissemination.
The fraction of citations from the destination to the origin country, meant to capture baseline citation activity, is only weakly associated with diffusion ($\beta = 0.06$, 95\% $CI = [0.06, 0.07]$, odds ratio $\approx 1.07$), reinforcing the distinction between general citation flows and meaningful idea uptake.

In Model (2), we incorporate additional structural and cultural variables, including topical distance, geographic distance, and shared language. 
Topic distance is negatively associated with idea diffusion ($\beta = -0.35$, 95\% $CI = [-0.35, -0.34]$, odds ratio $\approx 0.71$), indicating that ideas are more likely to spread between countries with thematically similar research agendas. 
Geographic distance emerges as a very modest but statistically significant deterrent ($\beta = -0.01$, 95\% $CI = [-0.01, 0.0]$, odds ratio $\approx 0.99$), suggesting that physical proximity still plays a role, albeit limited, in shaping transnational knowledge flows. 
Finally, sharing a common language increases the likelihood of diffusion ($\beta = 0.04$, 95\% $CI = [0.03, 0.05]$, odds ratio $\approx 1.05$), likely by reducing linguistic and cultural barriers to accessing foreign research; however, this effect loses significance once network-based features are included.

{\renewcommand{\arraystretch}{0.7} 
\begin{table}[H] \centering 
    \small
  \caption{\textbf{International diffusion of scientific ideas.}
Model coefficients for a series of weighted logistic regression models to predict the probability of ideas diffusing from an origin country to a destination country. The dependent variable is the observed fraction of ideas diffused from the origin country to the destination country, and the weights correspond to the total number of ideas originating from the origin country. Confidence intervals in parentheses. Standard errors and p-values are reported. McFadden's pseudo-$R^2$ is reported.}  
  \label{table:idea_diffusion}
\begin{adjustbox}{max width=\textwidth} 
\begin{tabular}{lllll}\\[-1.8ex]\hline 
\hline \\[-1.8ex] 
\multicolumn{5}{c}{\textbf{Dependent variable: Fraction of publications in top venues.}} \\[0.8ex]\hline 
 & \multicolumn{4}{c}{Model}  \\ \cline{2-5} \\[-1.8ex] & (1)& (2)& (3)& (4) \\ \hline \\[-1.8ex] 
Intercept & $-1.21^{***}$ & $-1.33^{***}$ & $-1.35^{***}$ & $-1.35^{***}$\\ 
 & $(-1.23,-1.2)$ & $(-1.34,-1.31)$ & $(-1.36,-1.33)$ & $(-1.37,-1.34)$\\ 
 & S.E. $0.01$; p-v $0.0$ & S.E. $0.01$; p-v $0.0$ & S.E. $0.01$; p-v $0.0$ & S.E. $0.01$; p-v $0.0$\\ [0.8ex]  
log GDP per capita origin & $0.35^{***}$ & $0.3^{***}$ & $0.29^{***}$ & $0.29^{***}$\\ 
 & $(0.34,0.35)$ & $(0.29,0.31)$ & $(0.28,0.3)$ & $(0.28,0.3)$\\ 
 & S.E. $0.0$; p-v $0.0$ & S.E. $0.0$; p-v $0.0$ & S.E. $0.0$; p-v $0.0$ & S.E. $0.0$; p-v $0.0$\\ [0.8ex]  
log GDP per capita destination & $0.19^{***}$ & $0.11^{***}$ & $0.1^{***}$ & $0.08^{***}$\\ 
 & $(0.18,0.19)$ & $(0.11,0.12)$ & $(0.09,0.1)$ & $(0.08,0.09)$\\ 
 & S.E. $0.0$; p-v $0.0$ & S.E. $0.0$; p-v $0.0$ & S.E. $0.0$; p-v $0.0$ & S.E. $0.0$; p-v $0.0$\\ [0.8ex]  
log Number of publications origin & $-0.36^{***}$ & $-0.33^{***}$ & $-0.3^{***}$ & $-0.3^{***}$\\ 
 & $(-0.36,-0.35)$ & $(-0.34,-0.32)$ & $(-0.31,-0.29)$ & $(-0.31,-0.29)$\\ 
 & S.E. $0.0$; p-v $0.0$ & S.E. $0.0$; p-v $0.0$ & S.E. $0.01$; p-v $0.0$ & S.E. $0.01$; p-v $0.0$\\ [0.8ex]  
log Number of publications destination & $1.66^{***}$ & $1.6^{***}$ & $1.58^{***}$ & $1.56^{***}$\\ 
 & $(1.65,1.66)$ & $(1.59,1.6)$ & $(1.57,1.59)$ & $(1.56,1.57)$\\ 
 & S.E. $0.0$; p-v $0.0$ & S.E. $0.0$; p-v $0.0$ & S.E. $0.0$; p-v $0.0$ & S.E. $0.0$; p-v $0.0$\\ [0.8ex]  
Fraction of citations to origin & $0.06^{***}$ & $0.06^{***}$ & $0.05^{***}$ & $0.05^{***}$\\ 
 & $(0.06,0.07)$ & $(0.06,0.06)$ & $(0.05,0.05)$ & $(0.05,0.06)$\\ 
 & S.E. $0.0$; p-v $0.0$ & S.E. $0.0$; p-v $0.0$ & S.E. $0.0$; p-v $0.0$ & S.E. $0.0$; p-v $0.0$\\ [0.8ex]  
Topic distance &  & $-0.35^{***}$ & $-0.34^{***}$ & $-0.33^{***}$\\ 
 &  & $(-0.35,-0.34)$ & $(-0.35,-0.34)$ & $(-0.34,-0.33)$\\ 
 &  & S.E. $0.0$; p-v $0.0$ & S.E. $0.0$; p-v $0.0$ & S.E. $0.0$; p-v $0.0$\\ [0.8ex]  
log Physical distance &  & $-0.01^{*}$ & $0.01^{***}$ & $0.01^{***}$\\ 
 &  & $(-0.01,-0.0)$ & $(0.01,0.02)$ & $(0.01,0.02)$\\ 
 &  & S.E. $0.0$; p-v $0.0104$ & S.E. $0.0$; p-v $0.0$ & S.E. $0.0$; p-v $0.0$\\ [0.8ex]  
Common language &  & $0.04^{***}$ & $0.02^{***}$ & $0.0$\\ 
 &  & $(0.03,0.05)$ & $(0.01,0.03)$ & $(-0.01,0.01)$\\ 
 &  & S.E. $0.01$; p-v $0.0$ & S.E. $0.01$; p-v $0.0$ & S.E. $0.01$; p-v $0.9992$\\ [0.8ex]  
Positive citation preference &  &  & $0.17^{***}$ & $0.13^{***}$\\ 
 &  &  & $(0.15,0.18)$ & $(0.12,0.14)$\\ 
 &  &  & S.E. $0.01$; p-v $0.0$ & S.E. $0.01$; p-v $0.0$\\ [0.8ex]  
Negative citation preference &  &  & $-0.03^{***}$ & $-0.04^{***}$\\ 
 &  &  & $(-0.04,-0.02)$ & $(-0.05,-0.03)$\\ 
 &  &  & S.E. $0.0$; p-v $0.0$ & S.E. $0.0$; p-v $0.0$\\ [0.8ex]  
Same network community &  &  &  & $0.08^{***}$\\ 
 &  &  &  & $(0.07,0.09)$\\ 
 &  &  &  & S.E. $0.01$; p-v $0.0$\\ [0.8ex]  
Network centrality origin &  &  &  & $0.01^{***}$\\ 
 &  &  &  & $(0.01,0.02)$\\ 
 &  &  &  & S.E. $0.0$; p-v $0.0$\\ [0.8ex]  
Network centrality destination &  &  &  & $0.04^{***}$\\ 
 &  &  &  & $(0.04,0.05)$\\ 
 &  &  &  & S.E. $0.0$; p-v $0.0$\\ [0.8ex]  
\hline 
\hline \\[-1.8ex] 
\textit{Note:} & \multicolumn{2}{r}{$^{*}p<0.05$; $^{**}p<0.01$; $^{***}p<0.001$} \\ 
Observations & 13837 & 12032 & 12032 & 12032 \\ 
pseudo-$R^2$ & 0.4109 & 0.4242 & 0.4246 & 0.4248 \\ 
Log Likelihood & -877425.86 & -799585.7 & -799065.57 & -798745.22 \\ 
F statistic & $145403.78^{***}$ (d.f.=5.0) & $84757.2^{***}$ (d.f.=8.0) & $67983.38^{***}$ (d.f.=10.0) & $52382.95^{***}$ (d.f.=13.0) \\ 
\hline 
\end{tabular}
\end{adjustbox}
\end{table}
}

The network of scientific recognition enhances our ability to predict the flow of ideas between countries, as shown in Models (3) and (4), with McFadden’s pseudo-$R^2$ values of $0.4246$ and $0.4248$, respectively (Table \ref{table:idea_diffusion}).
In Model 3, the presence of a positive citation preference between an origin and destination country is associated with a 1.18-fold increase in the odds of idea diffusion compared to the baseline of no significant preference ($\beta = 0.17$, 95\% $CI = [0.15, 0.18]$). 
Conversely, a negative citation preference corresponds to a 0.96-fold decrease in the odds of diffusion, indicating that countries that systematically under-recognize each other are less likely to exchange ideas ($\beta = -0.03$, 95\% $CI = [-0.04, -0.02]$). 
Adding an indicator for whether the origin and destination countries belong to the same recognition network community further improves the model (Model 4), showing a positive and statistically significant association with idea diffusion ($\beta = 0.08$, 95\% $CI = [0.07, 0.09]$). 
This suggests that countries embedded within the same broader recognition structure are more likely to exchange ideas, likely due to shared norms, collaborations, or visibility.
However, once community membership is included, the effect of the individual positive edge becomes attenuated, indicating that some of its predictive power is explained by broader structural cohesion within the network. 
This highlights the importance of both dyadic citation preferences and meso-level network positioning in shaping the global diffusion of scientific ideas.

Beyond the immediate neighborhood, the global network topology is hypothesized to play a significant role in the spread of information over social networks\cite{kempe2005influential, pei2018theories}.
Consistent with this view, Model (4) shows that both the centrality of the origin and destination countries in the recognition network are positively associated with idea diffusion. 
The centrality of the origin country has a modest but significant effect ($\beta = 0.01$, 95\% $CI = [0.01, 0.02]$), while the destination country’s centrality has a stronger association ($\beta = 0.04$, 95\% $CI = [0.04, 0.05]$). 
These findings suggest that ideas are more likely to spread from and to countries that occupy more prominent positions in the global recognition network, underscoring the importance of structural embeddedness in shaping transnational knowledge flows.

In addition, we perform a 5-fold cross-validation for Model (4) and evaluate its performance using McFadden’s pseudo $R^2$ (see SI, Table S11). The results indicate consistent model performance across different data partitions, suggesting that the model captures a substantial portion of the variation in the outcome and demonstrates reliable predictive power in out-of-sample contexts.

In summary, these findings emphasize that it is not just how much countries publish or cite, but how they are positioned and recognized within the global scientific system that influences the cross-national diffusion of ideas.

\section*{Discussion}

The international scientific landscape, a complex and dynamic web of knowledge, people and practices, is molded by national interests grounded in historical events, cultural values, political agendas, economics, and technological innovations.
These same forces shape interactions between nations through incentives for international collaboration, researcher mobility, and knowledge flows.
By analyzing more than fifty-seven million scientific publications across 223 countries spanning the period 1990-2022, we provide a large-scale temporal and structural analysis of the collective structure of global scientific recognition.  
We find that the international citation preference network constructed from these publications is shaped by cultural elements, including language and political agreements, and augments insights from the study of scientific collaboration and scientific topics.
Additionally, we quantify the network's departure from a core-periphery structure and identify five communities corresponding to major global regions, revealing a growing trend towards increased fragmentation.
Finally, we demonstrate that the international citation preference network captures constraints on the dissemination of scientific ideas, reflecting a more efficient spread of concepts within a community compared to their transmission between distinct communities.

Our analysis reveals the collective structure of international citation preferences, offering a novel perspective that complements prior work focused on collaboration, researcher mobility, and citation volume~\cite{leydesdorff2008internationalcollab, wagner_open_2017, glanzel_national_2001}.
By quantifying both the magnitude and statistical significance of national citation preferences, we provide a robust empirical foundation for assessing patterns of global scientific recognition.
However, our data do not permit identification of the full set of causal mechanisms driving these preferences.
Additional work is needed to disentangle the roles of cultural proximity, language, accessibility, and systemic bias. 
Still, the resulting network model of global recognition highlights important and underexplored dimensions of how nations acknowledge each other's contributions to science.

Our findings reveal a clear temporal evolution in the structure of global scientific recognition.
In the early 1990s, the citation preference network closely resembled a core-periphery model, with countries like the United States occupying a dominant central position and exerting broad influence across scientific communities.
This configuration reflected a relatively unified system in which recognition flowed predominantly from peripheral to core countries. 
However, beginning in the 2000s, this structure begins to fragment, giving way to the emergence of distinct regional clusters.
Rather than observing a smooth transition of countries from periphery to core, we find that some regions are becoming more internally cohesive yet increasingly disconnected from other parts of the network.
This shift is most clearly evidenced by the presence of negative citation links between communities, signaling declining mutual recognition---a pattern that would remain invisible in standard citation or collaboration networks.
While this pattern of growing fragmentation may indicate a shift toward internal knowledge exchange over global engagement, it may also be the result of long-standing preferences becoming more visible due to the exponential growth of global publication volume and the widening scope of the citation network.
Regardless of the underlying cause, the implications are significant: rather than converging toward a unified system of scientific influence, global science may be fracturing into regional communities that prioritize internal recognition at the cost of broader visibility and integration.
This dynamic risks deepening global inequalities, as previously peripheral regions become more insular, complicating efforts to foster inclusive and equitable recognition in science.

Our results further show that the global scientific landscape is not simply defined by a single core-periphery hierarchy but increasingly characterized by the emergence of distinct regional communities.

These findings deepen ongoing conversations in the sociology of science.
Scientific recognition, long understood as a gatekeeping mechanism, operates through socially embedded evaluative practices that determine whose knowledge is legitimized and whose contributions are marginalized~\cite{lamont2009professors, miao2024persistenthier}
Our evidence of persistent regional clustering in citation preferences and the limited cross-community diffusion of ideas echoes Cozzens’\cite{cozzens1989citations} insight that citation is not merely intellectual credit but a social signal. 
Similarly, our work supports Connell’s\cite{connell2014using} critique of global ``knowledge formations'' that privilege dominant scientific centers while marginalizing others.
At the same time, by revealing how regional communities amplify social structures and cultural contexts while inhibiting external exchange, our results challenge the Mertonian ideal of universalism~\cite{merton1973sociology} and highlight the fragmented nature of contemporary scientific recognition.

Our findings also carry important implications for science policy in an increasingly fragmented global research landscape. 
Overreliance on traditional core-periphery frameworks and global citation metrics risks overlooking vibrant, locally impactful science that does not align with established centers.
By revealing distinct, internally cohesive communities with differing levels of international engagement, our results suggest the need to diversify benchmarks of research impact. 
Policymakers and institutions should diversify evaluation benchmarks to recognize influence within regional or transregional contexts. 
Investments in horizontal collaboration, regional infrastructure, and locally relevant publication platforms can help mitigate fragmentation and promote more inclusive scientific integration.

At the same time, the presence of persistent negative citation preferences between communities underscores the urgency of rethinking how we fund and design cross-national partnerships. 
As shown in prior work~\cite{hoffman_international_2022}, bilateral agreements do not always translate into mutual recognition. 
Targeted efforts to build equitable and reciprocal scientific relationships---especially across underrepresented regions---are essential for addressing structural asymmetries and building a more globally integrated and equitable scientific community.


\section*{Materials and Methods}
\label{sec:methods}

\subsection*{Bibliometric Data}

The dataset was drawn from the OpenAlex\cite{priem2022openalex} bibliometric database in July 2022.  
OpenAlex is built upon the Microsoft Academic Graph (MAG), which was shuttered by Microsoft in December 2021, CrossRef, and ORCID.
We used all indexed ``journal-article'' and ``proceedings-article'' records listed as published between 1990 and 2022, and excluded any publication that did not list an institutional address, resulting in 76,080,360 publications.
We must acknowledge that the OpenAlex database has known limitations, including incomplete affiliation coverage~\cite{zhang2024missing} and a primary focus on English-language journals, which may introduce a selection bias towards Western countries~\cite{gong2019citationadv}. 
Despite these constraints, our results effectively identify significant patterns in scientific recognition.

Publications are associated with countries using the institutional addresses listed by the authors.  
We assign a full unit credit of a publication to every country of affiliation on the paper's author byline (``full counting'').
For example, a paper listing ten authors-- three with affiliations in Hungary, five with affiliations in the United States, and two in Canada--- would count one paper to all three countries.
See Supplementary Information for more details.

To account for potential sources of bias in national citation preferences, we implemented additional filtering steps to refine our citation dataset. 
First, self-citations---citations between publications authored by the same individual---are known to vary systematically across cultures, disciplines, and demographic groups, and may artificially inflate national recognition measures\cite{aksnes2003macro, King2017genderselfcite, azoulay2020selfcite}. 
Second, prior work shows that shared institutional affiliations can increase the likelihood of citation due to geographic or organizational proximity\cite{wuestman2019geography}. 
To mitigate the influence of these factors, we removed all citation links between publications that share at least one author (removing 88,078,384 citations) or at least one institutional affiliation (removing 20,769,688 citations).  
For additional details, see SI, Section S2.

\subsection*{National Co-variate Data}

We use data on national GDP per capita and Population from the World Bank\cite{fantom2016worldbank} to approximate each country's economic wealth and size.  
The dataset covers $264$ countries from 1960 to 2023.
The official spoken language is provided for $195$ countries and is encoded as a binary variable denoting common language for country pairs \cite{melitz2014native}.
We also source the bilateral distances (in kilometres) for most country pairs across the world from the \textit{GeoDist} dataset provided by the Centre for Prospective Studies and International Information (CEPII)\cite{mayer2011notes}. 
This dataset also provides the continent each country belongs to, which we convert into a binary indicator denoting whether two countries belong to the same continent. 
In addition, Science and Technology Agreements (STA) are regarded as an important tool to achieve strategic Science Diplomacy (SD) objectives \cite{langenhove2017tools}. 
We select records of STAs between countries \cite{nicolas2017case} to obtain the cumulative number of STAs between two countries over time.

\subsection*{National citation preference}

We fix a year $y$ and a source country (citing country) $s$ and identify all publications with at least one affiliation in the source country over the next $5$ years ($y$ to $y+5$).
We then find all publications worldwide published in year $y$ that also received citations from the source country's 5-year publications. 
This process generates country-specific citation frequencies ($c_{s,5}$) over the five-year observation window, enabling us to establish a hierarchical ranking of $n_{s,y}$ publications that have garnered at least one citation from the source country ($c_{s,5}>=1$).
This forms the baseline sample, comprising a citation distribution $p(c_5|s,y)$ specific to the source country $s$ and year $y$, with a sample size of $n_{s,y}$. 
Next, we narrow our analytical focus to a designated target country (cited country) $t$ , identifying a subset of $n_{s,t,y}$ publications within our sample $n_{s,y}$. 
These publications, represented by the distribution $p(c_5|s,t,y)$, must satisfy two criteria: they have received citations from the source country $s$ and maintain at least one institutional affiliation within the target country $t$.

The national citation preference, $P_{s,t,y}$, from the source country $s$ to the target country $t$ in year $y$ is found using the Area Under the receiver-operator Curve (AUC) as a measure of the extent to which the target country's publications are randomly distributed throughout the source country's ranking.
Specifically, the national preference is found as:
\begin{equation}
P_{s,t,y} = \frac{1}{n_{s,t,y} n_{s,y} } \sum_{i=1}^{n_{s,t,y}} \sum_{j=1}^{n_{s,y}} \mathbb{I}\left(c_{y,5}^{(i)} > c_{y,5}^{(j)}\right)
\end{equation}
where $c_{y,5}^{(i)}$ is the $i$-th sample from $p(c_5|s,t,y)$, $c_{y,5}^{(j)}$ is the $j$-th sample from $p(c_5|s,y)$, and $\mathbb{I}$ is the indicator function, which is 1 if $c_{y,5}^{(i)} > c_{y,5}^{(j)}$ and 0 otherwise.
The AUC is a measure of the probability (between 0 and 1) that a randomly chosen publication from the cited country is ranked higher than a randomly chosen publication from any other country; a value of 1 reflects the cited country's publications are over-expressed towards the top of the ranking, 0 occurs when the cited country's publications are under-expressed towards the bottom of the ranking, and 0.5 denotes a random distribution throughout the ranking.

We can further quantify the statistical significance of the over/under-representation of a specific country in the citation counts due to the equivalence of the AUC and Mann-Whitney U statistic (a.k.a. the Wilcox rank sum statistic).
Specifically, we follow DeLong et al. to compare the observed AUC to 0.5\cite{DeLong1988areaauc} using the algorithm's fast implementation \cite{Sun2014fastauc}.

\subsection*{International citation preference network}

The international citation preference network is a temporal network, with a snapshot layer generated for each publication year.  
To ensure statistical reliability, we include only country pairs with at least 50 cited publications in the relevant year, providing a sufficient number of data points for a robust preference to be expressed.
To correct for multiple hypothesis testing across all country pairs in a given year, we used the Holm step-down method\cite{holm1979simple} using Bonferroni adjustments as implemented in Statsmodels with $\alpha=0.01$.
Finally, we visualize the cumulative network which aggregates all yearly slices and assigns each edge the sign from the most recent year in which it appeared, reflecting the latest observed preference in the historical window.

The community structure within the positive international citation preference network is found using the Degree Corrected Stochastic Block Model (DCSBM) as implemented in graphtool\cite{peixoto2017nonparametric}.
Network centrality for the positive international citation preference network is found using the PageRank algorithm with a return probability of $\alpha=0.85$.
Finally, we characterize the inequality in the distribution of PageRank centrality across nodes using normalized entropy---a value between 0 and 1, where 1 indicates a perfectly uniform distribution (all nodes are equal) and 0 reflects extreme inequality, with influence concentrated in a small number of nodes.

\subsection*{Stratified bootstrap baseline}

To account for potential explanatory factors such as disciplinary focus and journal quality, we refine the assumptions underlying the random baseline in our national citation preference measure.
We achieve this by implementing a stratified bootstrap approach, where we sample from the conditional citation distribution while ensuring that the sampled set exactly matches the observed publication counts for each journal in the observed citation distribution. 
Specifically, given the sample of $n_{s,t,y}$ publications affiliated with the target country $t$ in year $y$ and cited by the source country $s$, we track the frequency with which each journal appears, denoted $j_{s,t,y}$.
We then sample with replacement from the source country's baseline distribution $p(c_5|s,y)$ such that the journal counts remain consistent with the observed values.
This adjustment controls for the influence of journal-specific factors and disciplinary differences.
We then perform 100 samples of this bootstrap procedure and use the mean and standard deviation of the AUCs to identify statistically significant links.

\subsection*{Scientific ideas}

To identify scientific ideas, we follow the methodology introduced in Cheng et al.\ 2023 \cite{cheng_how_2023}.
Specifically, we analyze the titles and abstracts for all of the publications in our OpenAlex corpus to identify the publications that mention at least one of 46,535 scientific ideas derived by Cheng et al. using the data-driven phrase segmentation algorithm, AutoPhrase \cite{shang2018automated}. 
We then post-process these ideas, removing cases that were first mentioned before 1990 and focusing only on those ideas that were mentioned by only one country in their first year of usage, resulting in 22,413 unique ideas mentioned in 752,075 publications.
Finally, we derive a dyadic variable for all pairs of countries in our network denoting the fraction of ideas whose first usage was in the Origin country and then were later used by a publication in the Destination country.

\subsection*{Weighted logistic regression analysis}

We use a weighted logistic regression model to the relationship between the propensity for scientific ideas to diffuse between countries and their connectivity in the international citation preference network.
The model is written as follows:
\begin{equation}
\log\frac{y_{c}}{1 - y_{c}} = \beta_0 + \beta_1X_{1c} + \beta_2X_{2c} + \ldots + \beta_kX_{kc}
\end{equation}
\noindent where $c$ denotes countries and $y_{c}$ is the probability that an idea originating in one country is eventually mentioned in the destination country.
In the first set of models, the dependent variable is the fraction of ideas originating in the source country that are subsequently mentioned by the destination country. 
We apply weights based on the total number of ideas produced by the source country, ensuring that countries with larger pools of originating ideas contribute proportionally to the estimation. (See Methods and SI, Section S5 for details.).
The included control variables are the GDP per capita and Number of Publications for both the Origin and Destination countries.
The investigated independent variables are the Topical Distance between the countries' publications, the Physical Distance between the countries, a binary indicator of common official language, the one-hot encoding of a directed positive edge from the Destination to the Origin in the international citation preference network, the one-hot encoding of a directed negative edge from the Destination to the Origin in the international citation network, and the PageRank centrality of the Origin and Destination countries in the positive international citation preference network.
We apply log-transformation with base 10 to GDP per capita, Number of Publications, and Physical Distance.
All features besides the binary features (Same Official Lang, Positive Edge, Negative Edge) are standardized by subtracting the mean and dividing by the standard deviation.

\subsection*{Fixed-effect multinomial logistic regression}

We use the multinomial logit model to predict the trinary citation preference between countries (e.g. positive, negative, or no preference). 
The multinomial logit model assumes that the log odds of each category $s\in\{-1,1\}$ relative to the reference category of no citation preference ($s=0$) is a linear combination of the independent variables. 
Specifically, the model is defined as follows:
\begin{equation}
\log \left(\frac{P(Y_{ijt}=s)}{P(Y_{ijt}= 0)}\right) = \beta_{s0}+ \beta_{s1} X_{it}+\beta_{s2} X_{jt}+\beta_{s3} X_{ijt} + \alpha_{t}
\end{equation}
\noindent where $P(Y_{ijt}=s)$ is the probability of the edge sign between source country $i$ and target country $j$ at time $t$ taking value $s\in\{-1,1\}$; $X_{it}$ and $X_{jt}$ capture potential country-specific characteristics in the country $i$ and $j$ at time $t$, respectively, while $X_{ijt}$ represents potential pair-specific barriers or catalysts between country $i$ and $j$ at time $t$; $\alpha_t$ are the time-specific effects (intercepts) that capture the heterogeneity across time periods. 
$\beta_{s0}$ is the intercept for category $s$; $\beta_{s1}, \beta_{s2} $ and $ \beta_{s3}$ are the coefficients associated with the independent variables $X_i, X_j$ and $X_{ijt}$ for category $s$. 
We investigate different variants of the above model to study different combinations of country-specific and country-pair-specific variables.
The included control variables are the GDP per capita, population, and the fraction of top journal publications for both the Source and Target countries.
The investigated pair-specific independent variables are physical distance, field distance, the same continent, the same official language, the cumulative number of bilateral science and technology agreements and scientific collaboration strength. 
We apply log-transformation with a base $10$ to GDP per capita, population, physical distance, the cumulative number of bilateral science and technology agreements and scientific collaboration strength.
All non-binary features are standarized by subtracting the mean and dividing by the standard deviation.

\clearpage 

%
\bibliography{natbiasref} 


\section*{Acknowledgments}
Special thanks to Albert-Laszlo Barabasi and the wonderful research communities at Northeastern's Center for Complex Network Research and UVA's School of Data Science for helpful discussions.  
We also thank Y.-Y. Ahn, Caroline Wagner, and Dashun Wang for helpful discussions and comments.

\paragraph*{Funding:}
A.J.G. was funded in part by the National Security Data \& Policy Institute, Contracting Activity \#2024-24070100001. 

\paragraph*{Author contributions:}
A.J.G. designed the project and wrote the manuscript.  A.J.G., I.M. and J.G. conducted the analysis.  All authors reviewed the manuscript.

\paragraph*{Competing interests:}
There are no competing interests to declare.

\paragraph*{Data and materials availability:}
The primary dataset, OpenAlex, is freely available online at \url{https://openalex.org/}.
All code used to conduct the analysis and generate the figures, as well as the processed data and network structure, is included as part of the pySciSci Python package\cite{gates2023pyscisci}: \url{https://github.com/SciSciCollective/pyscisci/globalscience}.


\subsection*{Supplementary materials}
Materials and Methods\\
Supplementary Text\\
Figures S1 to S14\\
Tables S1 to S11\\
References \textit{(1-34)}\\ 


\end{document}